\documentclass[aps,pre,epsfig,color]{revtex4}
\usepackage{graphicx}
\usepackage{color}
\begin{document}
\def\be{\begin{equation}}
\def\ee{\end{equation}}     
\def\bfi{\begin{figure}}
\def\efi{\end{figure}}
\def\bea{\begin{eqnarray}}
\def\eea{\end{eqnarray}}

\title{Topological regulation of activation barriers on fractal substrates}

\author{Raffaella Burioni}
\affiliation{Dipartimento di Fisica e Scienza della Terra, and INFN, Gruppo Collegato di Parma, Universit\`a di Parma, Parco Area delle Scienze 7/A, I-423100 Parma, Italy.}

\author{Federico Corberi}
\affiliation {Dipartimento di Fisica ``E.~R. Caianiello'', and INFN, Gruppo Collegato di Salerno, and CNISM, Unit\`a di Salerno,Universit\`a  di Salerno, 
via Ponte don Melillo, 84084 Fisciano (SA), Italy.}

\author{Alessandro Vezzani}
\affiliation{Centro S3, CNR-Istituto di Nanoscienze, Via Campi 213A, 41125 Modena Italy, and Dipartimento di Fisica Scienza della Terra, Universit\`a di Parma, Parco Area delle Scienze 7/A, I-43100 Parma, Italy.}

\begin{abstract}
We study phase-ordering dynamics 
of a ferromagnetic system with a scalar order-parameter on fractal 
graphs. We propose a scaling approach, inspired by
renormalization group ideas, where a crossover between distinct
dynamical behaviors is induced by the presence of a length
$\lambda $ associated to the topological properties of the
graph. The transition between the early and the asymptotic 
stage is observed when the typical size $L(t)$ of the growing 
ordered domains reaches the crossover length $\lambda $. We consider
two classes of inhomogeneous 
substrates, with different  activated processes, where the effects of the free energy barriers can
be analytically controlled during the evolution. On finitely ramified graphs the free energy
barriers encountered by domains walls grow logarithmically 
with $L(t)$ while they
increase as a power-law on all the other structures.
This produces different asymptotic growth laws (power-laws
vs logarithmic) and different dependence of the crossover length
$\lambda $ on the model parameters. Our theoretical picture agrees very well with
extensive numerical simulations.

\end{abstract} 

\maketitle

\section{Introduction}

The slow relaxation of glasses, spin-glasses and phase-separating systems
is a subject of  paramount importance in non-equilibrium statistical mechanics.
In the case of homogeneous ferromagnets, i.e non-disordered magnetic systems
defined on homogenous substrates such as lattices,  the basic features of the dynamics
after a quench below the critical temperature are well understood in 
terms of the domain growth mechanism \cite{Bray94}. The unbounded growth
of the size of ordered domains with nearly equilibrium composition involve a scaling
symmetry due to the presence of a dominant length scale, which is manifested
in the lack of time-translational invariance and aging. Although this simple paradigm of
slow relaxation is expected to encompass a broad variety of situations, 
its applicability to more complex systems, such as spin-glasses, where
both homogeneity and ferromagnetism are spoiled, remains a debated issue \cite{Bouchaud97}.
An intermediate class of systems is that of non-homogeneous ferromagnetic 
materials \cite{Puri04}. In these systems space homogeneity is wrecked 
by spacial modulations of some relevant parameter which,   
in order to maintain the
system ferromagnetic, must coexist with the low-temperature order.
In this case phase-ordering kinetics is preserved, although interesting
new features may arise. 
The agent whereby homogeneity is spoiled may be random, like 
random fields, random bonds, or
dilution, or deterministic, as in the case of models defined on
deterministic graphs.  The kinetics of these systems have been an active area of research for quite some time now. An unifying feature is the ubiquitous
appearance of energetic barriers slowing down the motion of domains boundaries.
This happens because, due to space inhomogeneity, interfaces are pinned in
particular positions, and this may have dramatic consequences on the growth law and
on the properties of  correlation 
functions \cite{Henkel06,Henkel08,Lippiello10,Park10}.
Despite many experimental and theoretical studies \cite{Puri04}, a number of issues 
are still open. 
In particular, the conditions under which inhomogeneities 
may radically modify the asymptotic dynamics are not 
{\it a priori} known \cite{Lai88}.
Indeed there are cases
where disorder merely change the time units,
like the one-dimensional Ising model with random 
bond
 \cite{7old}, and others, like the same model but with disorder in the form of a random-field \cite{dec},
where profound qualitative modifications occur. For this second
class of systems the nature of the dynamical scaling symmetry and of the asymptotic growth law is not yet fully characterized.
Related to that,
the conjecture of a superuniversal behavior, namely the idea that scaling functions are robust with respect to the presence on inhomogeneities which do not to change the low temperature properties of the system \cite{Cugliandolo10}, has been proposed. 

Recently, in the context of disordered ferromagnets, the observed behavior
has been interpreted \cite{noirf} in terms of a picture inspired by renormalization group ideas where disorder introduces, next to the domains size, another characteristic length which
gives rise to a crossover pattern. In this paper we apply 
similar ideas to the realm of ferromagnets defined on non-homogeneous  physical \cite{review} fractal graphs,
which turn out to be a simple but paradigmatic case where energy barriers thoroughly modify the asymptotic behavior of phase ordering dynamics.
Indeed, at variance with the case of disordered systems, the nature of the 
energetic barriers and their scaling properties can be reasonably well understood and
they can be related to the topological inhomogeneity. This allows
one to make precise predictions on the growth laws, on the scaling
properties and on the crossover phenomena. These predictions conform quite well
to the outcome of numerical simulations of the dynamics of the Ising model
on graphs. The main result of this paper is the existence of a crossover from an early power-law behavior to a slower asymptotic growth. The latter may be power-law (with a smaller,
temperature dependent exponent) or logarithmic according to the structure of the barriers, which in turn depend on the topology of the graph. In particular, we conjecture power-laws to be associated to graphs
which sustain ferromagnetic order only at zero temperature, 
while logarithms are expected if an ordered phase exists.  The scaling functions of
correlation functions are sensitive to the crossover.  

This paper is organized as follows. In Sec. \ref{scalteo}, we provide an overview of 
domain growth laws and of the scaling framework  proposed in \cite{noirf} for 
disordered systems. In Sec. \ref{scalgraph} we specify and adapt the scenario to the case of ferromagnets on graphs, providing also
predictions for several quantities such as crossover lengths and
exponents. 
 In Sec. \ref{numeric}, we present numerical results for the phase-ordering of the Ising model 
 on three model fractal graphs, the Sierpinski carpet (SC), the Sierpinski gasket (SG), and the T-fractal (TF). For the first structure a low-temperature ordered phase exists, whereas it is not sustained in
 the other two. We will argue this fact to represent a basic difference
 for the dynamical evolution.
 The numerical results are interpreted using the scaling ideas of Secs. \ref{scalteo},\ref{scalgraph} and agree quite nicely with the predictions obtained there. 
 In Sec. \ref{concl}, we conclude the paper with a summary and a discussion on the generality of our results.

\section{Growth laws and dynamical scalings}\label{scalteo}

Domain growth kinetics is characterized by an ever increasing typical domain size
$L(t)$ after the quench of a system to a final temperature $T$ in the ordered phase. 
When $L(t)$ becomes the dominant length in the problem, dynamical scaling is observed, 
meaning that all other lengths
can be measured in units of $L(t)$. 
Such property is reflected for instance in the two-point/two-time order parameter 
correlation function ${\cal G}(r;t,t_w)$, where $r$ is the distance between the two
points and $t>t_w$, which in homogeneous systems scales as \cite{Furukawa89}
\be
{\cal G}(r;t,t_w)=\widehat {\cal G} [r/L(t),L(t_w)/L(t)],
\label{scaling}
\ee
$\widehat {\cal G}$ being a scaling function.

Generally speaking, the growth law and other properties such
as $\widehat {\cal G}$
may depend on several factors, e.g.  
temperature, conservation laws, dimensionality, order parameter symmetry, 
disorder and topology of the substrate. In the simplest cases, as in homogeneous
magnets with non-conserved dynamics, the kinetics proceeds without activation
events. This means that no free energy barriers are encountered in the evolution
and, consequently, phase-ordering is observed down to temperature $T=0$. 
In this case the typical domains size increases as $L(t)\sim t^{1/2}$. 
On the other hand, the kinetics of other coarsening systems require thermal activation.
The simplest example are perhaps homogeneous magnets with conserved dynamics.
In this case energy barriers arise as due to  
the microscopic evaporation-condensation mechanism underlying the evolution \cite{cop}.
In general, barriers can have different origin and scaling properties.
In non-homogeneous systems, like the disordered ones or those embedded on a 
non-homogeneous substrate, they are typically due to the pinning of interfaces
in certain preferred positions. 
In a wide class of disordered systems,
such as the Ising model with random fields \cite{noirf,dec,altrirf,altrirf2}, random bonds \cite{noirb},
or random dilution \cite{Park10}, this pinning effect slows down 
asymptotically the growth law to a logarithmic
form
\be
L(t)\propto [\ln(t/\tau)]^{1/\psi},
\label{logl}
\ee
where $\tau $ and $\psi$ are model-dependent quantities. 
With an Arrhenius form 
\be
t \propto   e^{{\cal E}/(k_BT)}
\label{arr}
\ee
for the time needed to escape a barrier with activation free energy ${\cal E}$, the problem of establishing the growth law is
directly related to the one of determining the typical height ${\cal E}$ of barriers at
a certain stage of the evolution. 
Indeed, inverting the relation (\ref{arr}) one concludes that the typical activation 
increases with the domains size as a power-law:
\be
{\cal E}(L)\propto  L^\psi.
\label{scalbarlog}
\ee
Eq. (\ref{scalbarlog}) is not the only possible situation,
since a different algebraic growth law 
\be
L(t)\propto t^{1/\zeta},
\label{zt}
\ee
with $\zeta$ a temperature dependent exponent,
is observed asymptotically in systems defined on a class of fractal structures \cite{noipowerlawscal,noipowerlawvec}
and pre-asymptotically in random field 
\cite{noirf} and random bond ferromagnets \cite{noirb,powerrb}.
Plugging Eq. (\ref{zt}) in Eq. (\ref{arr}) one arrives at 
\be
{\cal E}(L)\propto [\ln(L)^\zeta],
\label{scalbarpower} 
\ee
for the scaling of the barriers with the domains size.
A classification of phase-ordering systems according to the possible growth
laws is made in \cite{Lai88}.
In general, however, a complete characterization of $L(t)$ 
in the many varied instances of coarsening systems is lacking. 
Besides that, understanding the behavior 
of the scaling functions such as $\widehat {\cal G}$ in Eq. (\ref{scaling}) is also an open problem. 

In Ref. \cite{noirf}, it was proposed to unify the wide variety of behaviors observed in domain growth into a generalized scaling framework. The basic idea, 
originally conceived for 
disordered systems, is that the agent
responsible for inhomogeneity introduces an extra characteristic length $\lambda$.
In the case of  the Ising model with non-conserved dynamics defined on a inhomogeneous 
graph considered in
this paper (see section \ref{scalgraph}), the only parameters of the model are the ferromagnetic coupling 
constant $J$ and $T$, which enter in the combination $\epsilon=J/T$. 
Then it must be $\lambda = \lambda (\epsilon)$. The presence of $\lambda$ 
introduces a crossover phenomenon between two different dynamical behaviors
when $L(t) $ crosses $\lambda$.  For the domains size this is 
assumed to be described by
\be
L(t;\epsilon) = t^{1/z}\widehat L(\epsilon/t^\phi),
\label{cros}
\ee
where $z$ and $\phi $ are a growth and a crossover exponent, with a scaling function behaving as
\be
\left \{
\begin{array}{l}
\widehat L(x)=\mbox {const.,} \hspace{2cm} \mbox {for} \hspace {.5cm} x\to 0\\
\widehat L(x)=x^{1/(\phi z)}\ell (x^{-1/\phi}), \hspace{.5cm} \mbox {for} \hspace {.5cm} x\to \infty\\
\end{array}
\right .
\label{cross}
\ee
where $x = \epsilon/t^\phi$. 
In systems  where disorder plays the role of an irrelevant 
parameter $\phi$ is positive, inhomogenities only affects the dynamics at early times and $L(t)$ crosses over to the 
pure system behavior $L(t)\sim t^{1/z}$ at large times \cite{noirf}.
This usually happens if there is an upper limit for the height
of energetic barriers. On the other hand, when barriers of any size
can be encountered, as in the fractal models considered in this paper and 
others, $\phi$ is negative and 
Eq. (\ref{cross}) describes the crossover from the early stage power law
\be
L(t)\sim t^{1/z},
\label{lt}
\ee
with a temperature independent growth exponent $z$,
to the asymptotic form $L(t)= \lambda (\epsilon)\ell(t/\epsilon ^{1/\phi})$
at the crossover length $\lambda=\epsilon ^{1/(\phi z)}$.
The functions $\lambda(\epsilon)$ and $\ell(t/\epsilon ^{1/\phi})$ are in general hard to
be determined, both analytically and numerically. However, some prediction 
can be made in the case of coarsening on deterministic fractal structures, as will be discussed
in Sec. \ref{scalgraph}.

\subsection{Autocorrelation function}

In order to discuss the effect of inhomogeneities on the form
of the correlation functions let us consider as a paradigm 
the autocorrelation function 
$C(t,t_w)={\cal G}(r=0;t,t_w)$.
Generalizing the crossover approach (\ref{cros}) to
this two-time quantity, one would expect the form
\be
C(t,t_w;\epsilon)=\widehat C\left[\frac{L(t)}{L(t_w)},\frac{\lambda (\epsilon)}{L(t_w)}\right],
\label{autocscal}
\ee
where $\widehat C$ is a scaling function. In the context of disordered ferromagnetic systems 
a superuniversality conjecture was proposed according to which the effect of disorder is simply
accounted for by the slower growth of $L(t)$, while scaling functions entering correlation functions
remains unchanged with respect to the clean case. This amounts to say that, e.g. for the autocorrelation
function,  $\widehat C(x,y)$ should not depend on the second entry. The superuniversality property
was checked in several models, arriving at different conclusions. 
Indeed, while  $d = 1$ results \cite{dec,7old} clearly demonstrate the absence of 
superuniversality, in the cases with $d \ge 2$, there is evidence both in favor \cite{altrirf2,againstsu} and against \cite{Lippiello10,noirf} its validity, and there is presently 
an intense debate on the subject. 

The original formulation of superuniversality was conceived for systems
where inhomogeneities are introduced by disorder. The problem of determining the relevance of
inhomogeneities on the scaling functions, however, may be posed on more general grounds
and the simple systems studied in this paper may help the clarification of this issue.

\section{Coarsening on fractal structures} \label{scalgraph}

\begin{figure}[h]
    \centering
   \rotatebox{0}{\resizebox{.5\textwidth}{!}{\includegraphics{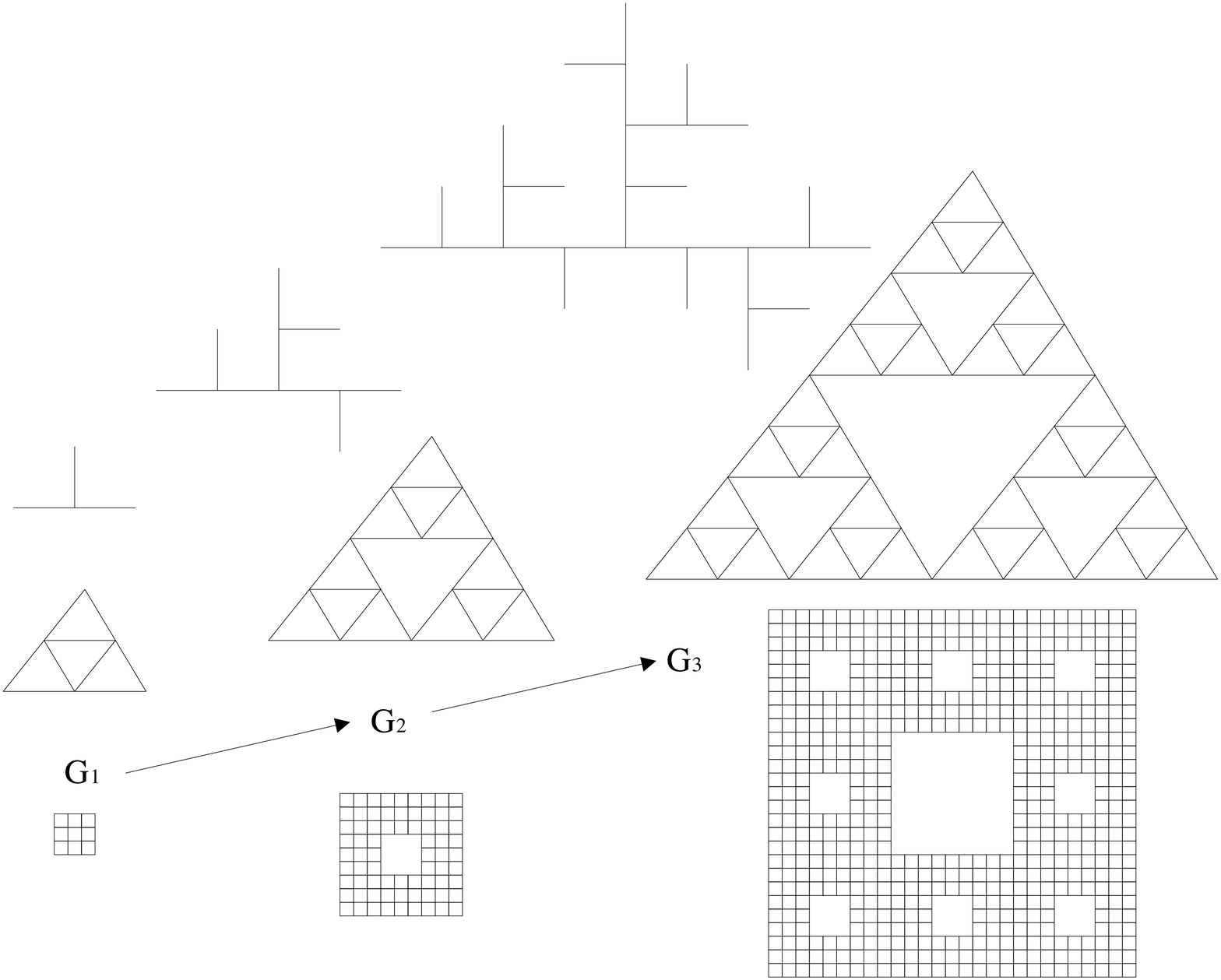}}}
   \vspace{1cm}
   \caption{Construction of the first 3 generations of the 
   TF (upper part), SG (central part) and of the SC (lower part).}
\label{fig_generations}
\end{figure}

We will consider in the following the Ising model
defined by the Hamiltonian 
\be
H[\sigma] = -J \sum _{\langle ij\rangle} \sigma_i \sigma_j,
\ee
where $\sigma _i= \pm 1$ is a unitary spin and $< ij >$ are nearest neighbors on a graph.
The dynamics is introduced by randomly choosing a single spin and updating it with a transition rate that in the numerical simulations will be chosen in the Metropolis form
\be
w([\sigma ]\to [\sigma ']) = \mbox{min} \{1, \exp[-\Delta E/(k_BT )]\} .
\label{metropolis}
\ee
Here $[\sigma ]$ and $[\sigma ']$ are the spin configurations before and after the move, and $\Delta E = H[\sigma '] - H[\sigma]$. Phase-ordering is observed after a quench from
an high initial temperature (assumed to be infinite in the following) to $T=0$
or to a temperature below the critical one $T_c$.
In order to infer the properties and the physical content of the quantities 
introduced in the previous section,
let us focus on the  structures of the SG, of the TF,  and of the SC \cite{gefen,noiprl}, representing  prototypical examples of finitely ramified 
(the first two) and non-finitely ramified fractals (the latter). Indeed a wide class of fractal structures feature topological properties similar to the SG, TF and SC \cite{frattalivari}.
In finitely ramified fractals, an arbitrary large part of the structure can be disconnected by removing a finite number of {\it cutting bonds}. In general, they admit an ordered phase only at $T=0$. In a broad sense these structures can 
be considered a non trivial topological analogous of a low dimensional homogeneous systems  with $d\le d_{\ell}$, $d_{\ell}$ being the lower critical
dimension. Non-finitely ramified graphs, on the other hand, can possess
a low temperature ordered phase below a critical temperature 
$T_c$.
All these fractals can be build starting from an elementary object, denoted as the first generation
$G_1$, obtaining then an object of generation $G_2$ by combining $G_1$ parts, and then proceeding
recursively as sketched in Fig. \ref{fig_generations}. In doing that, the linear size $L_n$ 
of the structure at generation $n$ grows as $L_n=f\cdot L_{n-1}$, where $f$ is constant which
depends on the structure considered and, specifically, $f=2$ 
for the SG and the TF, and $f=3$ for the SC.

Since the Ising model on finitely ramified fractals is characterized by $T_c=0$, phase-ordering could in principle  be truly asymptotic
only after a quench to $T=0$, where however the dynamics
is frozen because activated moves are prevented. 
Nevertheless, as explained in \cite{noipowerlawscal},
a preasymptotic coarsening stage of diverging duration  is observed for 
sufficiently low
temperatures, similarly to what happens for instance on the $1d$ lattice. 
We will focus on this stage of the dynamics to study coarsening
on finitely ramified structures.

\begin{figure}[h]
    \centering
   \rotatebox{0}{\resizebox{.5\textwidth}{!}{\includegraphics{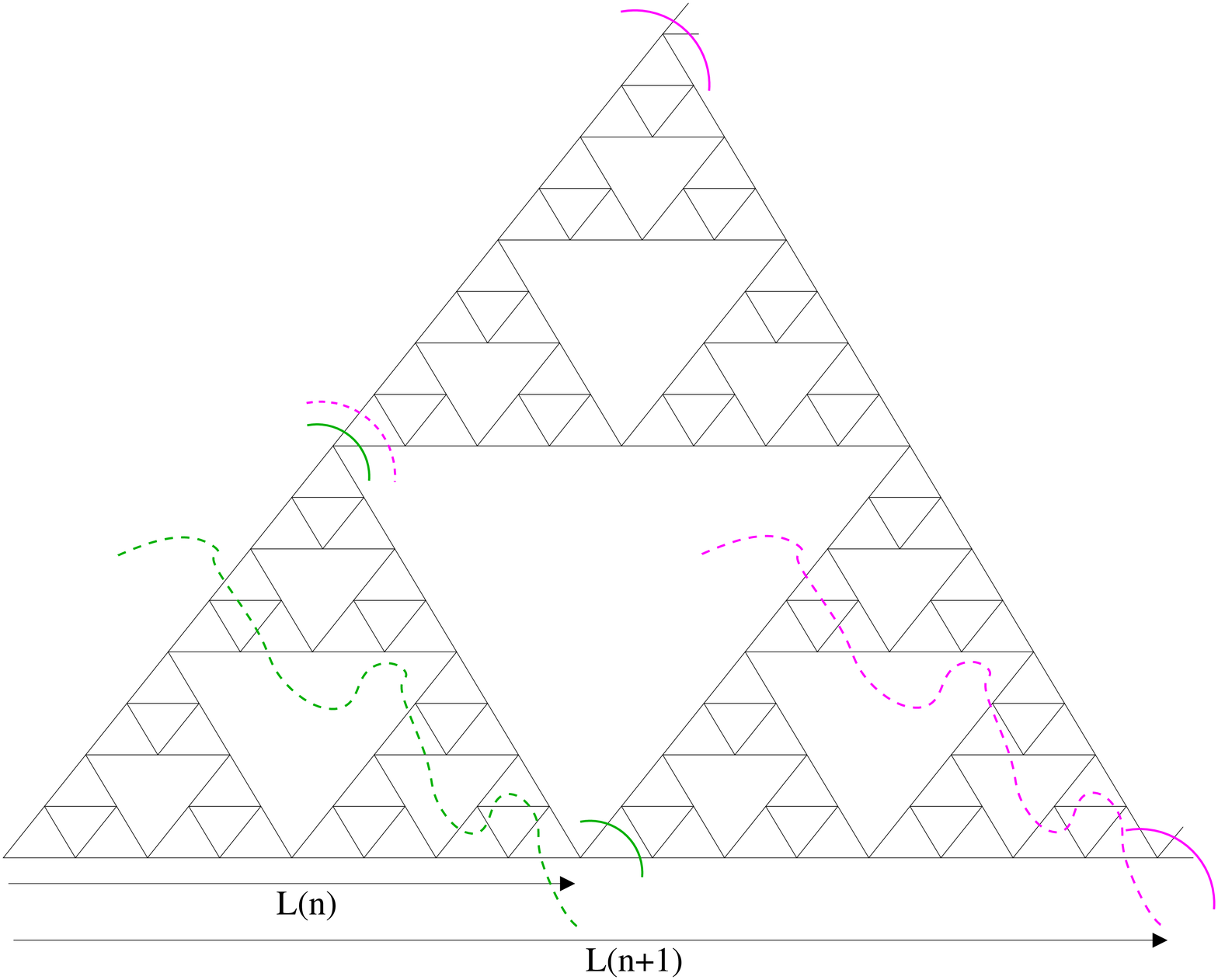}}}
   
   \vspace{1.5cm}
   
   \rotatebox{0}{\resizebox{.5\textwidth}{!}{\includegraphics{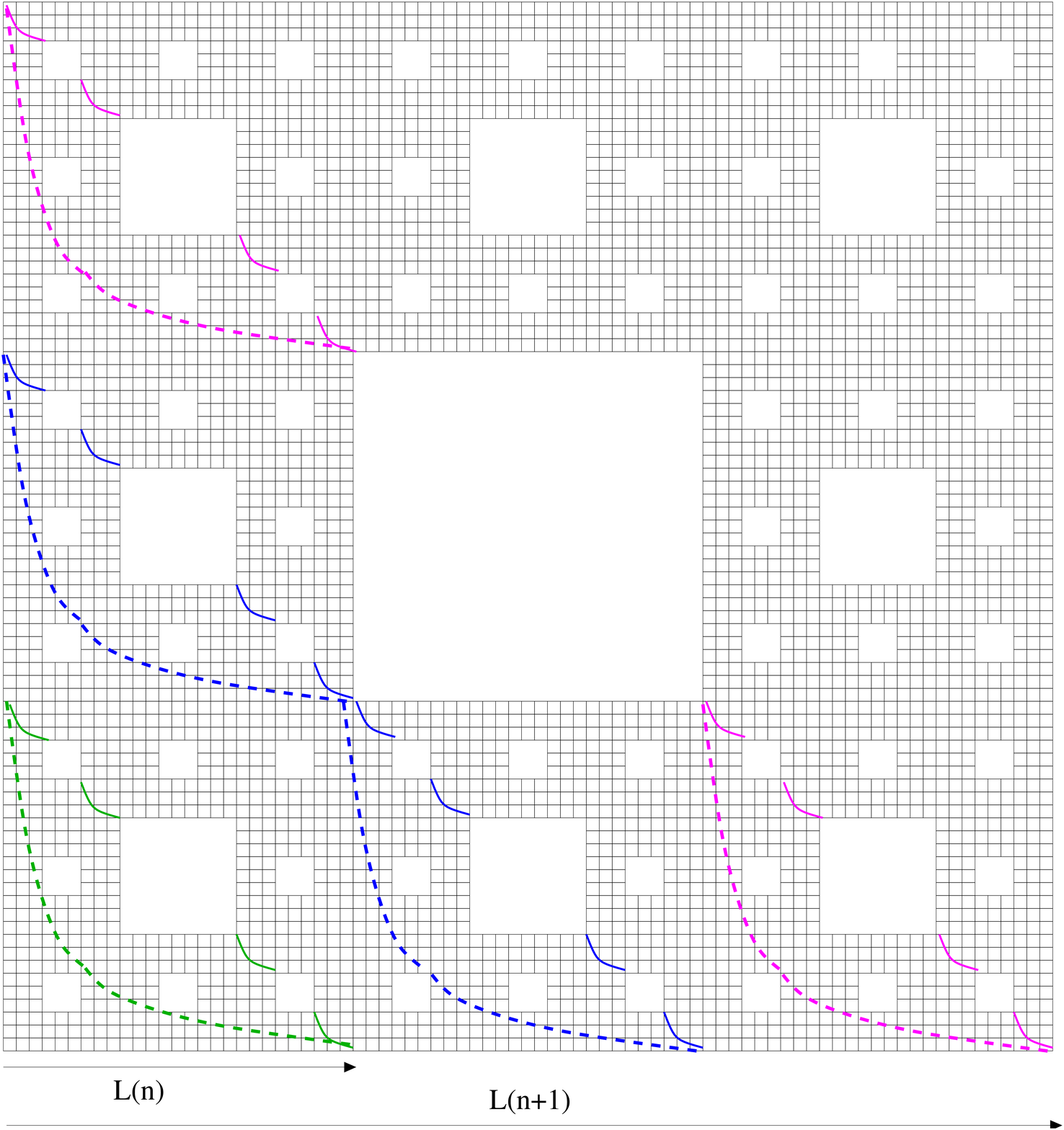}}}
   
   \vspace{1cm}
   
\caption{Schematic description of energy barriers in the SG 
(upper panel) and in the SC (lower panel). 
For the SG, the green and the magenta arches represent the position of the interfaces in a configuration of minimum energy
after a region of generation $n$ (and $n+1$) has been reversed. 
The dashed green and magenta lines represent the configuration of 
the interface with larger energy while reversing the structure of 
generation $n$ (and $n+1$).
For the SC, the continuous green, blue, and magenta lines represent 
the position of minimum energy of an interface after the the reversing
of a part of structure of generation $n$ (and $n+1$). 
The corresponding dashed lines are the position of larger energy to be exceeded.}
\label{fig_gasket}
\end{figure}

During the evolution, energetic barriers arise due to the mechanism sketched in Fig. \ref{fig_gasket}. 
Let us consider the SG first, represented in the upper panel.
The figure shows schematically the evolution of an interface
which progressively spans a part of the structure. Initially, the position of the domain wall 
is outside the structure represented in the figure, in the left corner. This means that
all the spins are, say, down.
As time goes on, the interface enters the graph by moving across the intermediate position
$I_n^{(max)}$ indicated by a dotted green line (this means that spins on the left of the green line 
have been reversed up). The index $n$ refers to the fact that the interface is currently spanning the $n$-th generation of the fractal.
When the spins of the whole generation $n$ have been reversed, the interface,
depicted with two green arches, is located in the configuration $I_n^{(min)}$ on the four {\it cutting bonds}. 
Since the energy of a domain wall is $J$ times its length, it is clear that
in $I_n^{(min)}$ the interface has a minimum energy $E_n^{(min)}$. On a finitely ramified structure, since the number of
{\it cutting bonds} is finite
this quantity is bounded from above and for large $n$, it becomes $n$-independent, i.e. $E^{(min)}_n=E^{(min)}$. For the SG, $E^{(min)}=4J$ for any $n$, since there are four
{\it cutting bonds} for the interface, independently of $n$.
Let us assume that the highest energy $E_n^{(max)}$ of the system during the above process was reached
in the (generic) configuration $I_n^{(max)}$, so that a barrier of height 
${\cal E}_n=E_n^{(max)}-E_n^{(min)}$ has been crossed.
Now the interface must proceed again to the right in order to reverse all the spins
of the next generation $n+1$, thus reaching the position $I_{n+1}^{(min)}$ located on the other four {\it cutting bonds},
and indicated by the couple of magenta arches (all the spins in the figure at this stage have then been reversed). 
Also this configuration has an energy equal to $E^{(min)}$.
The topology of the finitely ramified graph allows the system to 
attain this configuration by sequentially reversing parts of
generation $n$ of the structure. For instance, in the Sierpinski 
gasket, by first reversing another triangle of
generation $n$, say the lower-right one in Fig. \ref{fig_gasket}.
This event is analogous to the one described before.
In particular, the interface in the intermediate position $I_{n+1}^{(max)}$, depicted with a dashed-magenta line, 
is analogous to the previous one at $I_n^{(max)}$
(dotted-green), except for the presence of an extra part around certain cutting bonds which, in the present example, are indicated
with a dashed-magenta arch.
For low temperatures the number of these cutting bonds tends to be minimized and it
does not depend on the generation.
Denoting with $E_{n+1}^{(max)}$ the maximum energy reached by the
system in the reversal of the $n+1$ generation, 
one concludes that $E_{n+1}^{(max)}=E_n^{(max)}+E_c$, where $E_c$ is the amount of extra energy
due to the new part of interface (the dotted-magenta arch
in the figure). Writing $E_c=Jn_c$ where $n_c$ is the 
double of the number of
broken bonds associated to the energy $E_c$,
for the SG one has $n_c=4$. We recall that in general
$n_c$ does not depend on $n$. Then 
\be
{\cal E}_{n+1}\simeq {\cal E}_n+Jn_c,
\ee
or, equivalently, rewriting ${\cal E}$ in terms of the size $L_n$ of the $n$-th generation
\be
{\cal E}(f\cdot L_n)\simeq {\cal E}(L_n)+Jn_c.
\label{sgen}
\ee
From this relation, dropping the index $n$, one has
\be 
{\cal E}\simeq \frac{Jn_c}{\ln f}\ln L.
\label{sgel}
\ee
Inserting this form into the Arrhenius relation (\ref{arr}) we arrive at a algebraic growth law  as in Eq. (\ref{zt})
with
\be 
\zeta \simeq a\,\frac{J}{T}
\label{stimazeta}
\ee 
and 
\be
a\simeq \frac {n_c}{k_B\ln f} 
\label{stimaa}
\ee
at low temperatures. The argument has been developed 
referring to the deterministic SG but it is expected to hold 
also for the TF (with $n_c=2$) and in general for finitely ramified deterministic and disordered structures, as for all of them an arbitrary large part can be disconnected by cutting a finite number of links. 

A similar argument applies to non-finitely ramified structures, and it will be schematically illustrated
for the SC with the help of the lower panel of Fig. \ref{fig_gasket}.
As for the SG, an interface enters the portion of the fractal considered
from (say) the lower-left edge, 
crossing a region of generation $n$, passing through 
an high energy position depicted as an arch-shaped dashed green line.
The evolution then proceeds
towards the position $I^{(min)}_n$ which in the figure is depicted 
by a continuous 
green line that roughly spans the diagonal of the n-th generation. 
Here the energy reaches a minimum $E^{(min)}_n$, because the
largest hole available at generation $n$ is crossed, thus minimizing the number of disaligned spins.
Then the domain wall moves to the next high energy configuration, depicted as a couple of dashed arch-shaped blue lines, and then
to the minimum energy position $I^{(min)}_{n+1}$ depicted with a 
continuous diagonal blue line. 
Eventually the 
configurations magenta are progressively reached by
the interface. Notice that these configurations have the same
energy of the blue ones.
It is clear from the figure that the energies of
the blue (or magenta) configurations occupied at generation 
$n+1$ are the double of those of the green ones at generation $n$.
Hence one can write, in place of (\ref{sgen})
\be
{\cal E}(f\cdot L_n)=u{\cal E}(L_n),
\ee
where $u=2$ for the SC.
Hence, in place of Eqs. (\ref{sgel}) we find
\be
{\cal E}(L_n)\simeq bJL_n^{d_{fl}},
\label{sgel2}
\ee
where $d_{fl}=\ln u/\ln f$ is in general the fractal dimension of the intersection
of the fractal structure with a line, that is the border of the interface, corresponding to the cut set (e.g. it is $d_{fl}=\ln 2/\ln 3$ for the SC considered in the figure).   
and from Eq. (\ref{arr}) we arrive at Eq. (\ref{logl}),
with
\be 
\psi= d_{fl}.
\label{stimapsi}
\ee
This argument, developed for the SC, is expected again to be of general validity for all non-finitely ramified
structures.

In conclusion, according to the above discussion, finitely (SG and TF) and infinitely (SC) ramified fractals  represent the typical examples where barriers increase logarithmically and algebraically with $L$, Eqs. (\ref{scalbarlog}, \ref{scalbarpower}) respectively, due to their topological properties. Correspondingly
the growth law is expected to be a power law (with a temperature dependent exponent) or a logarithmic behavior, Eqs. (\ref{zt},\ref{logl}) respectively.
These behaviors should be observed when the energy scale $k_B T$ associated to 
temperature fluctuations is small with respect to the height ${\cal E}$ of the barriers. Conversely,
for ${\cal E}\ll k_BT$, the effect of the barriers is negligible.
Interfaces diffuse freely in this case and, in analogy
to what observed on regular lattices, one expects to observe a power law behavior as in Eq. (\ref{lt}), characterized 
by the temperature independent exponent 
which describes the displacement 
$\langle x^2\rangle \propto t^{1/z}$
of a random walker on the structure. For a fractal graph having 
spectral and fractal dimension $d_s, d_f$ respectively, it is
$z=2d_f/d_s$  \cite{fracrw}. 
We remark that this same power-law is predicted 
by approximate theories \cite{umb} to be the correct 
asymptotic growth-law, whereas our arguments indicate that it
can only be preasymptotic.  

All these behaviors can be fitted into the crossover scenario described by Eq. (\ref{cros})
provided that $z=d_s/(2d_f)$ and 
\be 
\left \{
\begin{array}{l}
\ell (x)\propto x^{1/\zeta } 
\hspace{1.7cm} ,\hspace{1cm} 
\mbox{for finitely ramified graphs}\\
\ell (x)\propto (\ln x)^{1/\psi } 
\hspace{1cm} ,\hspace{1cm} 
\mbox{for non-finitely ramified graphs,}
\end{array}
\right .
\label{growthlaw}
\ee
with $\zeta $ and $\psi $ given in 
Eqs. (\ref{stimazeta},\ref{stimapsi}).

The crossover length $\lambda $ is given by the condition
\be 
{\cal E}(\lambda )\simeq k_BT.
\ee
Using Eqs. (\ref{sgel}) and (\ref{sgel2}) this implies
\be 
\left \{
\begin{array}{l}
\lambda \sim \exp [k_BT\ln f/(Jn_c)]
\hspace{1cm} ,\hspace{1cm} 
\mbox{for finitely ramified graphs}\\
\lambda \sim \left ( k_BT/J\right )^{1/d_{fl}}
\hspace{2cm} ,\hspace{1cm}  
\mbox{for non-finitely ramified graphs}
\end{array}
\right .
\label{lambda}
\ee

\section{Numerical simulations} \label{numeric}

In this section we present the results of numerical simulations of the dynamics of the
Ising model after a quench on the SG, on the TF, and on the SC.
In the simulations we will consider systems with 
$n=9$, $n=8$ and $n=6$ generations respectively.
With these choices finite size effects are
not observed in the range of times considered. 
The system is prepared in a completely disordered state
where spins are set to $\sigma _i=\pm 1$ randomly and independently 
on each site $i$, corresponding to an infinite temperature
equilibrium initial state. The evolution is then implemented 
flipping randomly chosen single spins with the Metropolis rate
(\ref{metropolis}) where $T$ is the temperature at its quench value. 
We always
set $k_B=1$ and $J=1$ or, stated differently, we measure temperature
in units of $J/k_B$. We consider a modified dynamics called
no-bulk-flip, where spins which are aligned with all the nearest neighborg, namely in the bulk of an ordered domain, cannot flip. 
It was shown that this dynamics, which has been tested and
used in a number of studies \cite{noipowerlawscal,nobulk}, does not 
change the
large-scale properties of the system and improves the speed of the 
simulation and the quality of the results. In order to grasp at least the
basic physics inspiring the no-bulk-flip rule let us recall that 
in a coarsening system all the large-scale/long-time properties are
uniquely determined by the motion of the ordered domains walls.
Well inside the domains, regions of spins correlated over a length
$\xi (T)$ can rapidly change sign over a characteristic time 
$\xi (T)^z$, where $z$ is the dynamical exponent. However, since for large times $t$
and $T<T_c$, $\xi (T)$ is always negligible with respect to
$L(t)$, the flipping of these {\it thermal islands} only affect
small-distances/short-times properties which are not the universal ones
we are interested in. 
Further details 
on this accelerated dynamics can be found in 
\cite{noipowerlawscal,nobulk}.
For the SG, we study quenches to different final temperatures in the
range $T\in [0.8-3.4]$. For $T>3.4$ the system very rapidly reaches 
the disordered equilibrium state and the coarsening stage cannot be observed.
On the other hand, for $T<0.8$ the dynamics becomes exceedingly
slow to obtain reliable results with our computational resources.
Similarly, for the TF we consider quenches in range of final
temperatures $T\in [0.2-3]$  
and for the SC $T\in [1-3]$.

\begin{figure}[h]
    \centering
   \rotatebox{0}{\resizebox{.45\textwidth}{!}{\includegraphics{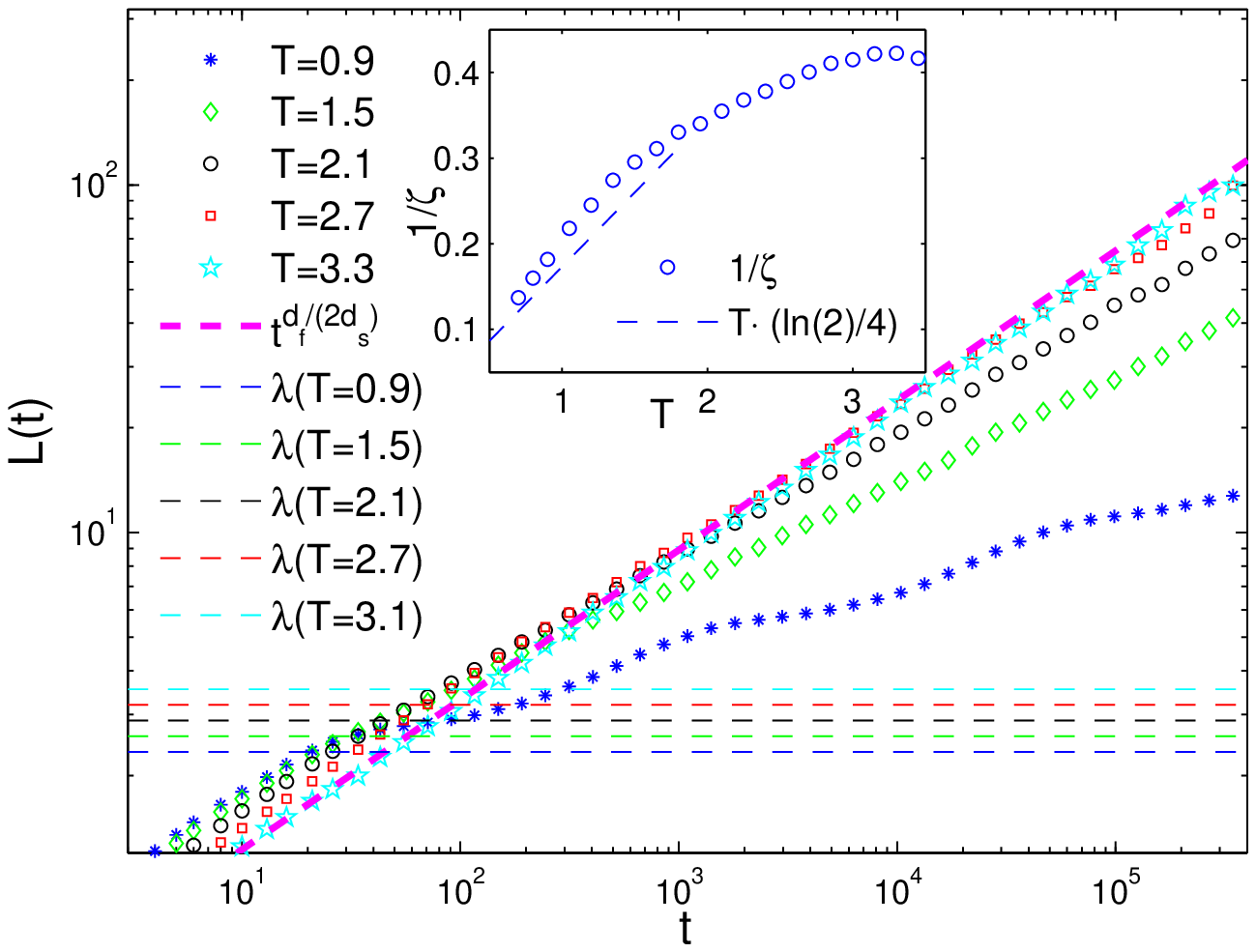}}}
   \rotatebox{0}{\resizebox{.45\textwidth}{!}{\includegraphics{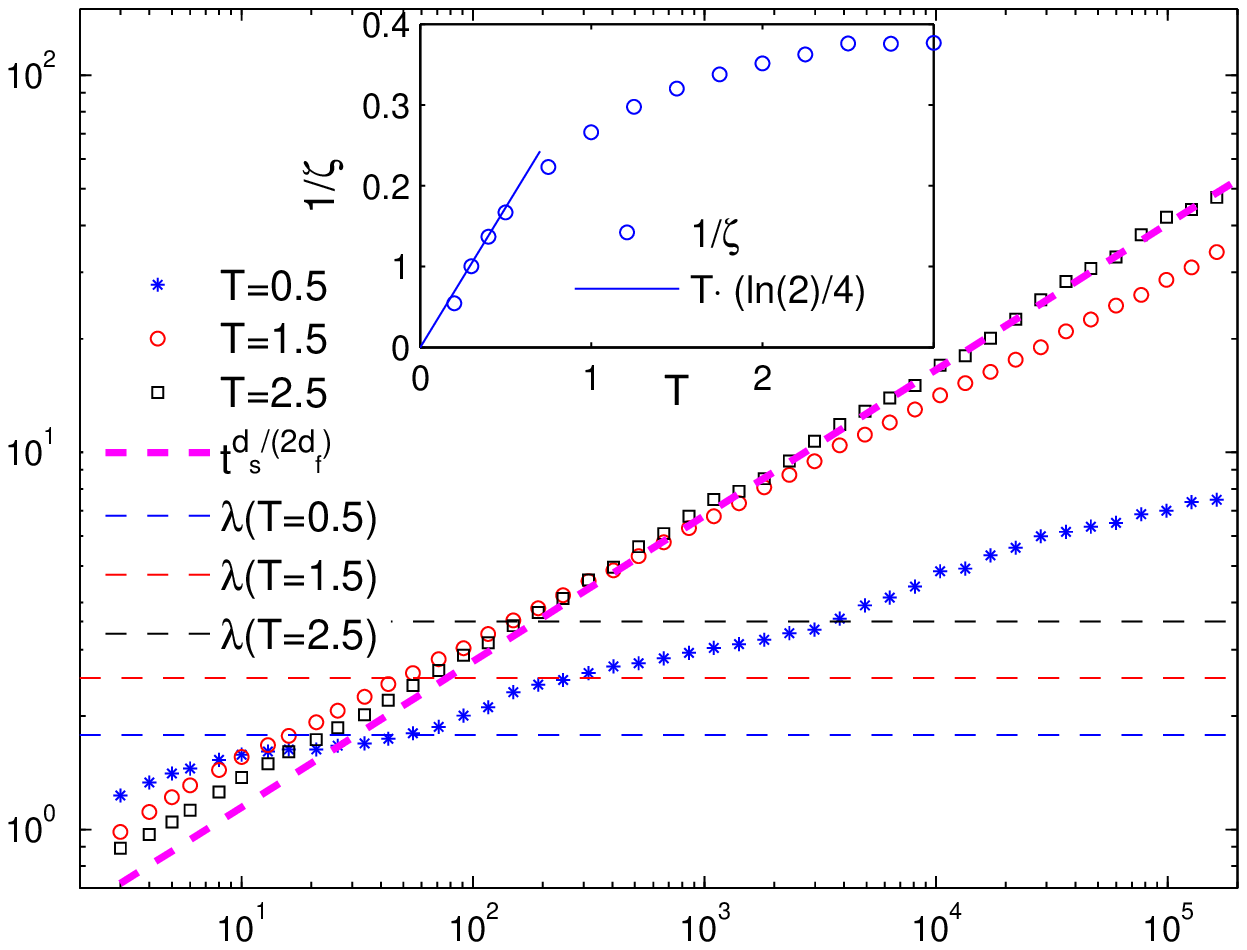}}}
   \rotatebox{0}{\resizebox{.45\textwidth}{!}{\includegraphics{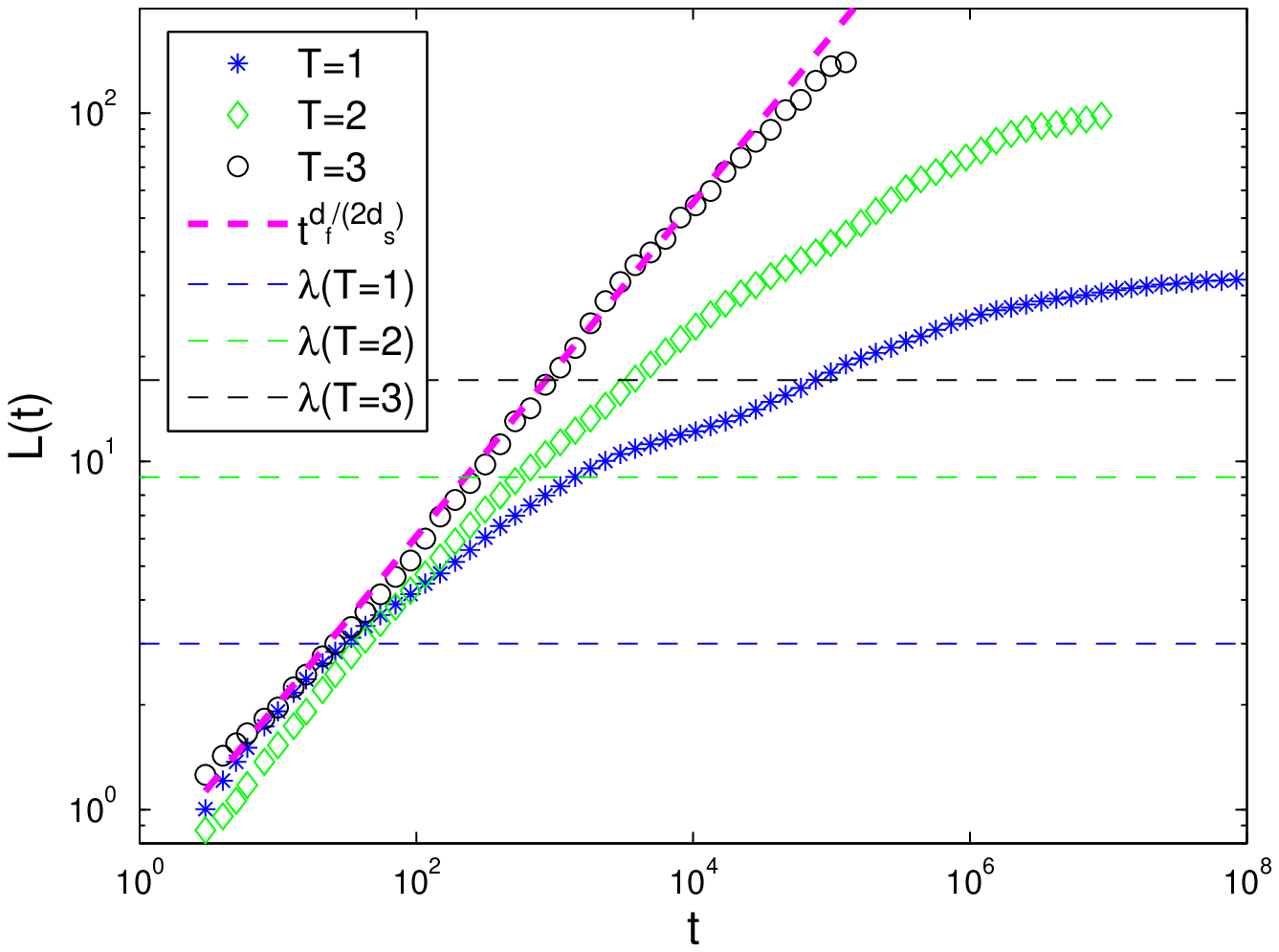}}}
   \rotatebox{0}{\resizebox{.45\textwidth}{!}{\includegraphics{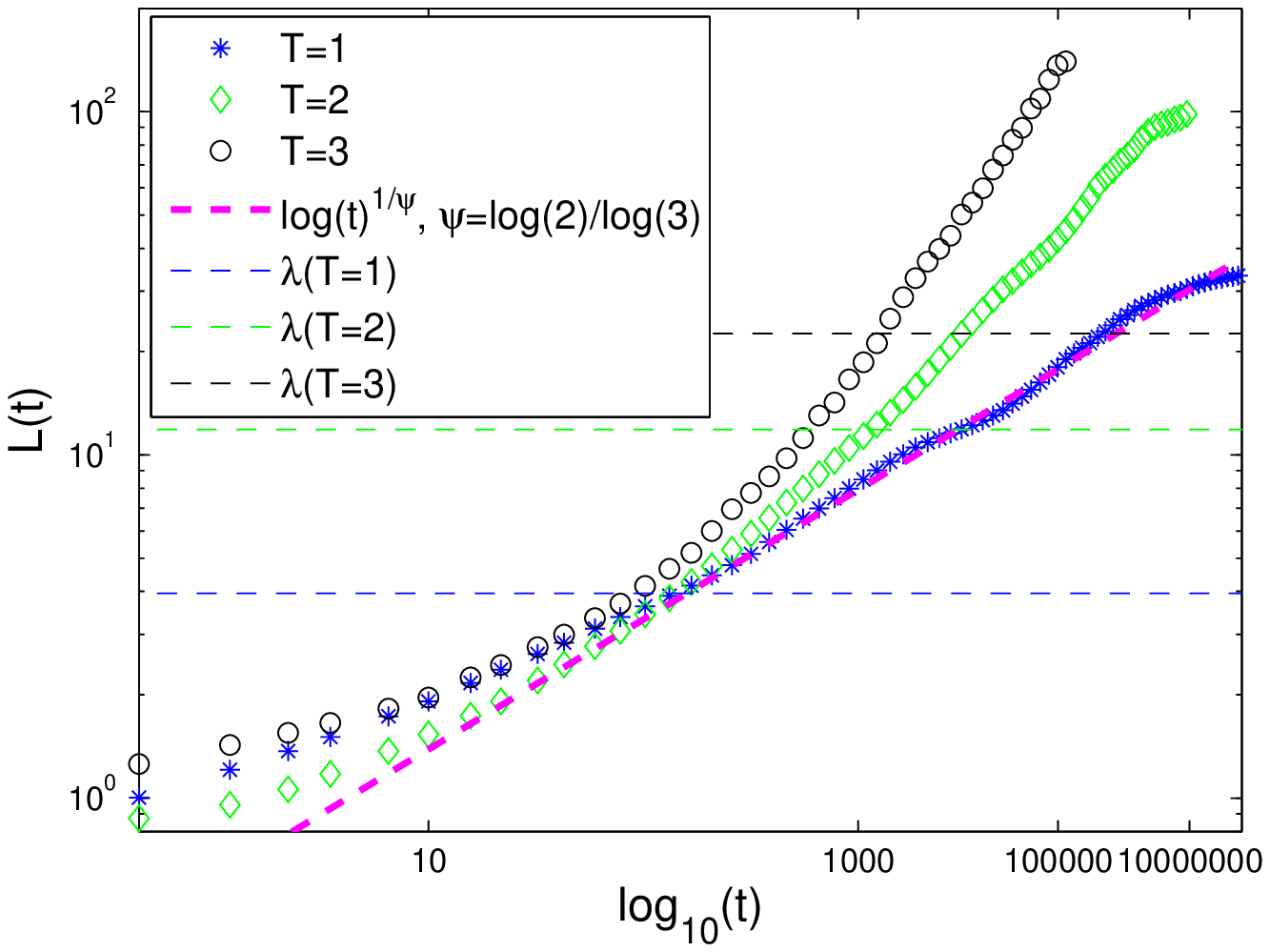}}}
   \vspace{1cm}
   \caption{The typical size $L(t)$ (symbols) is plotted against time in a log-log plot for an 
Ising model
quenched to different final temperatures (see key) on the SG (upper left panel), the TF (upper right panel) and on the SC (lower panels). The bold dashed
magenta line in the upper and lower left panels is the short time expected behavior $L(t)\sim t^{1/z}$, with $z= 2d_f/d_s$, while in the lower right panel
represents the asymptotic logarithmic law $L(t)\sim (\ln t)^{1/\psi}$,
with $\psi =d_{fl}=\ln 2/\ln 3$. The horizontal dashed lines represent the
crossover length $\lambda$.
In the inset of the upper panels the exponent $1/\zeta $, obtained from the 
data of the main part of the figure, is plotted against $T$.
The dashed blue line is the prediction (\ref{stimazeta},\ref{stimaa}).}
\label{fig_lengths}
\end{figure}

On a homogeneous structure, assuming scaling (\ref{scaling}), the equal time correlation function
\be
G(r;t)=\langle \sigma _i (t) \sigma _j (t) \rangle,
\ee
depends only on the distance $r$ between $i$ and $j$. 
From this the typical length $L(t)$ is usually extracted e.g. as the half height width 
$G(r=L(t);t)=1/2$.
Although on a generic graph the notion of distance is not straightforward, one can reduce its definition
to the usual one along certain directions, as done in 
\cite{noipowerlawscal,noipowerlawvec}. 
Proceeding analogously here we
arrive at a determination of $L(t)$.
This quantity is shown in Fig. \ref{fig_lengths} for quenches to different final temperatures. 

A first trivial observation is that in the structures 
considered in the simulations there is a lower cutoff $L_{hole}$ on the 
size of the {\it holes}. Hence for $L(t)<L_{hole}$ the fractal nature of the 
graph is
not revealed and we expect to observe the same behavior as on a 
homogeneous lattice. $L_{hole}$ is of the order of the size of the 
first generation and, in the cases considered here, $L_{hole}\simeq 3$.
Very early data with $L(t)<L_{hole}$, therefore, do not describe
the effect of the fractal structure and we will only refer the discussion
to $L(t)>L_{hole}$.

For all structures considered here one sees very clearly a crossover from an early regime 
$L(t)\ll \lambda$, where curves for different temperatures roughly collapse, to a late stage with a 
strongly temperature dependent behavior.   
This is what
one would expect from the picture described in the previous sections. 
According to the discussion of Sec. \ref{scalgraph}, 
if such a crossover is described by Eq.  (\ref{cros}),
a number of predictions can be done, that we will check
in the following. 

\vspace{.5cm}
{\bf Crossover length:}
\vspace{.2cm}

We start from the behavior of the 
crossover length $\lambda$ given in Eq. (\ref{lambda}).

For the SG,
using $f=2$ and $n_c=4$ one has $\ln f/(Jn_c)\simeq 0.17$. 
Using Eq. (\ref{lambda}) one concludes that, in the range of
temperatures explored ($T\in [0.8-3.4]$), $\lambda $ can be varied 
at most by a factor $1.57$.  In Fig. \ref{fig_lengths} 
(upper left panel) we have 
indicated with a dashed horizontal line the value of 
$\lambda$ obtained from Eq. (\ref{lambda}) by adjusting the
proportionality constant in such a way that, for $L(t)>\lambda$
only the asymptotic power-law growth (\ref{growthlaw}) (with
$\zeta \neq z$) is observed. This can be done rather precisely 
for the lowest temperatures. Since $\lambda $ can be varied
only by the small factor $1.57$ it is clear that, even for the highest
temperature one cannot obtain a significative range 
$L_{hole}<L(t)<\lambda$ to clearly determine the preasymptotic stage, 
where Eq. (\ref{lt}) should hold with a temperature-independent 
exponent $z=2d_f/d_s$ (this law is represented by the dashed magenta
line). Notice that at the highest temperatures $L(t)$ grows
approximatively as in Eq. (\ref{lt}) but, as will be discussed
below, for the different reason that the asymptotic 
temperature-dependent exponent $\zeta$ approaches the value
$z=2d_f/d_s$ at high temperatures.
A similar situation is found for the TF (upper right panel). 
Here the smaller value of  $n_c$ ($n_c=2$)
allows to vary $\lambda $ by slightly larger factor i.e. 2.46.
On this structure, then, at variance with 
the SG one should be able to detect the preasymptotic regime
$L_{hole}<L(t)<\lambda$ at least for the larger temperatures.

For the SC we have plotted in Fig. \ref{fig_lengths} (second and 
third panel)
the crossover length $\lambda $
obtained from Eq. (\ref{lambda}), by using $d_{fl}=\ln 2/\ln 3$
and adjusting the proportionality constant in such a way that  
the crossover from early power-law to asymptotic logarithmic
behavior occurs around $\lambda$.
As shown in Fig. \ref{fig_lengths} our estimate of $\lambda $
fit quite nicely at a semi-quantitative level with the data.

\vspace{.5cm}
{\bf Early stage:}
\vspace{.2cm}

Next, let us consider 
the first regime, for $L(t)\ll \lambda$, that should obey the
 power-law behavior (\ref{lt}) with the temperature independent
exponent $z=2d_f/d_s$. As already discussed above, this preasymptotic
regime is too short to be studied in the SG.
For the TF this should in principle be observed at least at the largest
temperature $T=2.5$
However, as it can be seen in Fig. \ref{fig_lengths}, 
the crossover appears to be
very broad, preventing a clearcut evidence on the preasymptotic
behavior also on this structure.

On the contrary, for the SC at the highest quench temperature $T=3$
this stage lasts for more than a decade.
Here one observes the expected power-law
with exponent $z=2d_f/d_s$ (such law is represented
with the dashed magenta line).
This shows that, indeed, in a preasymptotic regime barriers do not play any relevant
role and interfaces perform a random walk on the graph.
Notice also that, as already observed in \cite{noipowerlawscal,noipowerlawvec}, an oscillatory behavior is superimposed on a globally increasing trend.
This feature is observed in all the dynamical regimes 
and also in the SG and TF.
Although the very limited extent of the time that can be reached in
simulations does not allow to observe more than at most 1-2 oscillations,
thus preventing any precise analysis,
a semi-quantitative inspection of the data clearly suggests that these might
be log-time periodic. This periodicity is observed in a number of 
apparently different phenomena, ranging from fracturing of 
heterogeneous solids \cite{2627}, to stock market indexes \cite{28},
from magnetic systems with lack of translational symmetry 
\cite{andrade},
to phase separating fluids under shear \cite{cgl}. 
This feature is generally associated to the presence of
a discrete scale invariance \cite{sor} which in the present case induces
a recurrent trapping of
the interfaces when a complete generation of the fractal has been
ordered. The presence of such log-periodic
oscilations has been analytically proven for a random walker diffusing
on these fractal structures \cite{rw}.

\vspace{.5cm}
{\bf Late stage:}
\vspace{.2cm}

The next step is the determination of the asymptotic growth law, namely the function $\ell$, which should behave as in 
Eq. (\ref{growthlaw}).

For the SG and the TF it is clear that the curves oscillate around a net power-law behavior, as already
observed in \cite{noipowerlawscal}. 
This is what was expected for  finitely ramified graphs, 
according to Eq. (\ref{growthlaw}).
Our arguments provide the prediction (\ref{stimazeta},\ref{stimaa}) 
for the low-temperature behavior of $\zeta$. 
In order to check this we have plotted
$1/\zeta $ against 
temperature in the inset of the upper panels of 
Fig. \ref{fig_lengths}. For low T the data show a good agreement with 
the expected behavior, which is represented by the dashed-blue
lines. Notice that, in this case, there are no fitting parameters
since also the value of the constant $a$ has been inferred. 

In the case of the SC, the logarithmic growth of $L(t)$ forces
one to reach much longer times, particularly at low temperatures,
and this in turn increases the computational effort limiting
the possibility of a large statistics over many realizations of
the thermal histories. Moreover the screening due to the oscillations is 
more severe. Nevertheless the data of the second panel of
Fig \ref{fig_lengths} show quite unambiguously a 
crossover to a slower growth law around $L(t) \simeq \lambda$.
According to our general picture this should be described by the 
logarithmic form of Eq. (\ref{growthlaw}). In order to check this 
we have plotted in the third panel the same data but with an extra
logarithm on the time axis. In this plot the form of Eq. (\ref{growthlaw})
is a straight line with slope $1/\psi$. 
This is very well consistent with our data for $L(t)>\lambda$.
The numerical results show that the exponent $\psi$ decreases when $T$ is 
increased. In the limit of low temperature the value 
of Eq. (\ref{stimapsi}) is predicted. The corresponding law is
represented with a dashed-magenta line in the third panel of Fig.
\ref{fig_lengths}. Here one observes a very nice agreement with the
data for the lowest temperature ($T=1$), suggesting the correctness
of our argument.

\vspace{.5cm}
{\bf Autocorrelation function:}
\vspace{.2cm}

Finally, our simulations allows to comment on the role of the inhomogeneities on the scaling functions of
correlation functions. 
We will consider in the following the autocorrelation function
\be
C(t,t_w)=\frac{1}{N}\sum _{i=1}^N \langle \sigma _i(t)\sigma _i(t_w)\rangle,
\ee
where $N$ is the number of spins in the structure, for which
the scaling form (\ref{autocscal}) is expected.

\begin{figure}[h]
    \centering
   \rotatebox{0}{\resizebox{.5\textwidth}{!}{\includegraphics{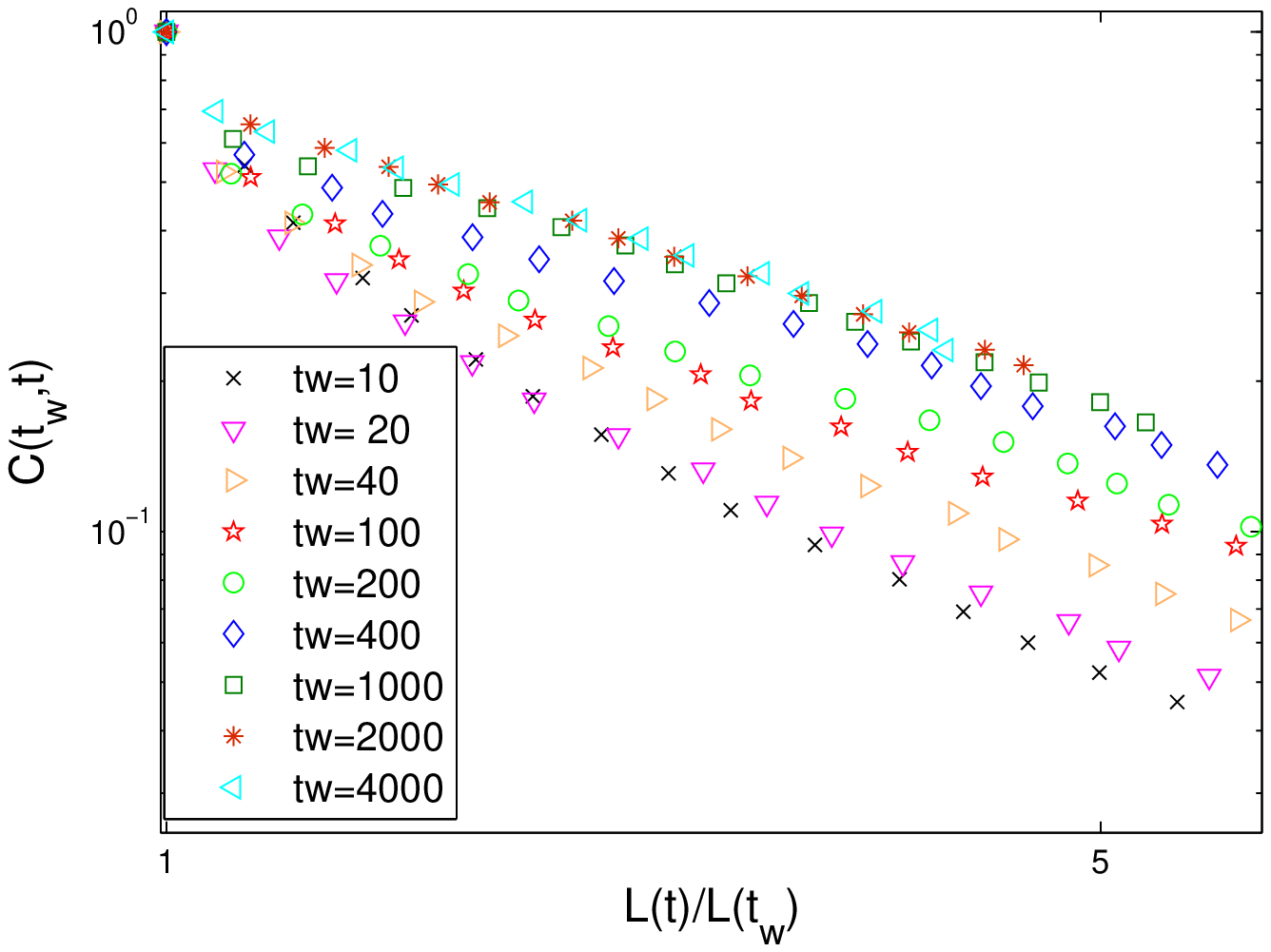}}}
   
   \vspace{1.5cm}
   
   \rotatebox{0}{\resizebox{.5\textwidth}{!}{\includegraphics{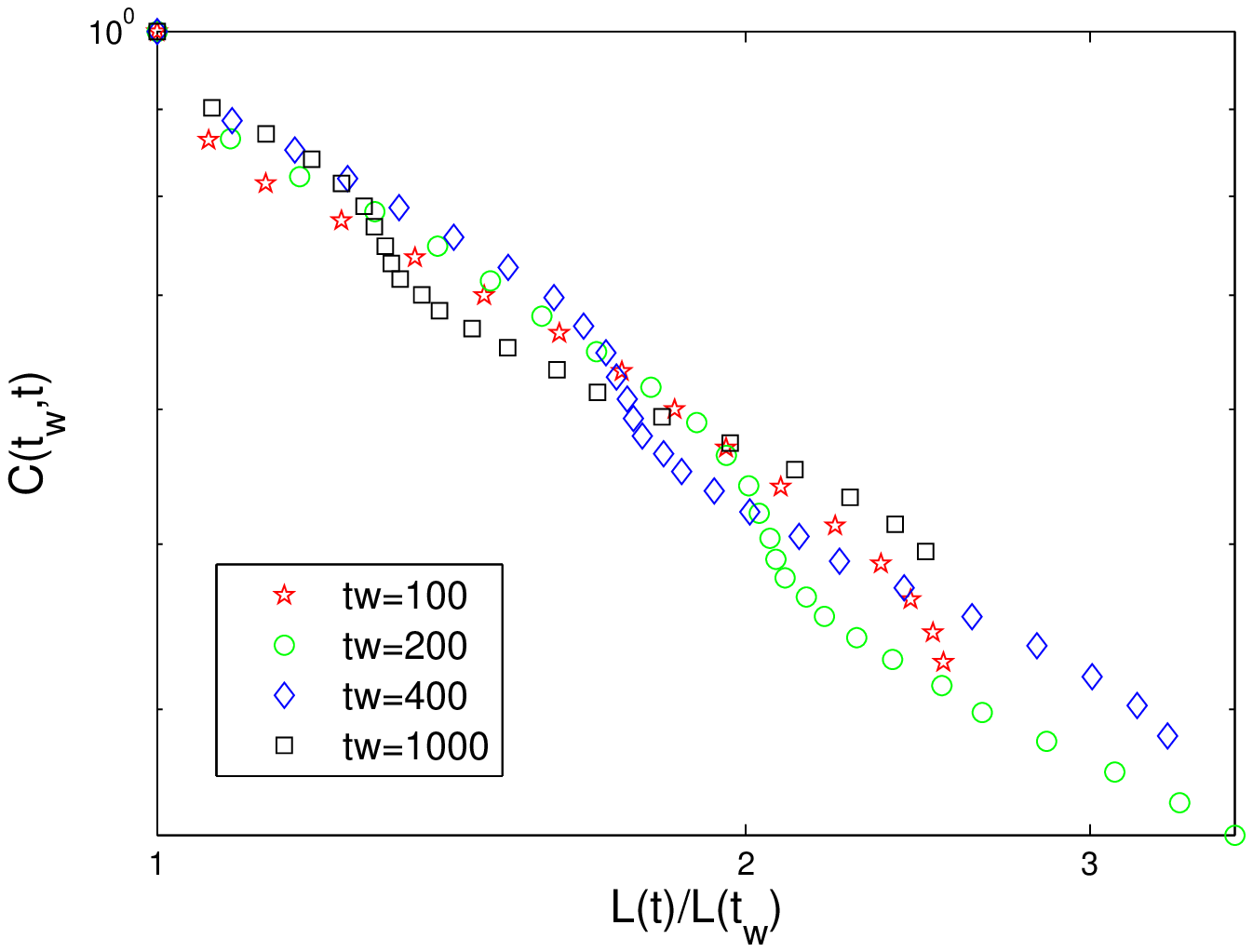}}}
   
   \vspace{1cm}
   
\caption{The autocorrelation function $C(t,t_w)$ is plotted against $L(t)/L(t_w)$
for different values (see key) of $t_w$. 
In the first panel a quench to $T=3$ is considered, while in the
second it is $T=1$.}

\label{fig_autocorr}
\end{figure}

As discussed previously, because of the smallness of $\lambda $
on the SG and of the broad form of the scaling functions in the TF, the whole crossover pattern can be better
detected in the SC. Hence we concentrate on this structure in the following.
In Fig. \ref{fig_autocorr} we plot the autocorrelation function measured
after different  waiting times $t_w$. 
The upper panel refers to the quench to the highest temperature 
$T=3$, where, according to the data of Fig. \ref{fig_lengths},
one can access both the preasymptotic regime and the crossover
to the late stage, and the role of the second entry of the scaling function
in Eq. (\ref{autocscal}) can be studied. 
Interestingly, the crossover phenomenon is fully displayed
in the figure. Indeed, the two curves for the smallest waiting times
($t_w=10$ and $t_w=20$) almost collapse. The small residual
dependence on $t_w$, namely the fact that the collapse is not
perfect, is probably due to the fact that the presence of the crossover
is already slightly felt even at these early times (but we cannot
reduce $t_w$ further because of the constraint $L(t_w)>L_{hole}$).
However, the picture shows that the separation between these
two curves is much smaller than the one among the following
ones, and this clearly indicates a convergence of the data to 
a limiting mastercurve for $L(t_w)\ll \lambda$. In view of 
Eq. (\ref{autocscal}) this means that the second entry 
$\lambda /L(t_w)$ is so large that 
$\widehat C\left [L(t)/L(t_w),\lambda (T)/L(t_w)\right ]
\simeq \widehat C\left [L(t)/L(t_w),\infty \right ]$.
Then, upon increasing $L(t_w)$, the second argument of 
$\widehat C$ decreases and becomes relevant. Indeed the collapse is lost and there
is a clear tendency of the curves to move to larger values,
as already noticed in \cite{noipowerlawscal}. 
This signals the crossover from the early stage to the late regime.
Finally, for larger values of $L(t_w)$
($L(t_w)\gg \lambda$) one has again a tendency to collapse on a 
mastercurve which 
corresponds to $\widehat C\left[ L(t)/L(t_w),0\right]$. 
This is very well verified for $t_w=1000, 2000, 4000$.

In order to 
complete the analysis we consider also the quench to
the lower temperature $T=1$. In this case, since $\lambda $ is smaller
the limiting curve $\widehat C\left [L(t)/L(t_w),0 \right ]$
should be achieved at earlier $t_w$. Our results are plotted
in the second panel of Fig. \ref{fig_autocorr}. 
At this low temperature the autocorrelation function
is strongly oscillating. This is the counterpart of the periodic
modulations already observed in the growth-law.
These oscillations hinder somehow the collapse of the curves.
Nevertheless, one observes that already from such small waiting
times as $t_w=100$ onwards the curves do not show any tendency
to move upwards, at variance with the case with $T=3$. This can be interpreted as due to the fact that,
since $\lambda $ is much smaller at $T=1$, the collapse on the mastercurve 
$\widehat C\left [L(t)/L(t_w),0 \right ]$ is already achieved at these
early times. Clearly, since $\lambda$ is small,
the pre-crossover collapse on $\widehat C\left [L(t)/L(t_w),\infty \right ]$ 
is not observed here. 
Notice that the two mastercurves $\widehat C\left [L(t)/L(t_w),\infty 
\right ]$ and $\widehat C\left [L(t)/L(t_w),0 \right ]$ in the first
panel are very well separated, clearly indicating the relevance of the 
second entry in the
scaling function of Eq. (\ref{autocscal}).
The whole behavior of $C$, which is captured by the 
two-parameter scaling (\ref{autocscal}), show unambiguously the 
relevance of the inhomogeneities,
entering through the length $\lambda$, in determining the shape of the 
scaling functions. 
 
\section{Conclusions and perspectives}\label{concl}

In this paper we have studied the phase-ordering kinetic 
of a ferromagnetic system with a scalar order-parameter on fractal 
graphs. We have proposed a scaling approach, inspired to
renormalization group ideas, where a crossover between distinct
dynamical behaviors is induced by the presence of a length
$\lambda $ introduced by the topological properties of the
graph. The transition between the early and the asymptotic 
stage is observed when the typical size $L(t)$ of the growing 
ordered domains reaches the crossover length $\lambda $.
In this general framework, two classes of inhomogeneous 
substrates can be defined according to the nature of the 
activated processes which set in during the evolution.
Specifically, we argue that on finitely ramified graphs the free energy
barriers encountered by domains walls grow logarithmically 
with $L(t)$ while they
increase as a power-law on all the other structures.
This produces different asymptotic growth laws (power-laws
vs logarithmic) and different dependence of the crossover length
$\lambda $ on the model parameters. We have tested these ideas 
by numerical simulations of the Ising model on two
model structures where, due to their relative simplicity, one 
can exhibit explicit predictions for the behavior of $L(t)$ and
of $\lambda$, which conform very well to the numerical data.

The models studied in this paper can be considered as simple 
prototypical systems to understand the more general and still
open problem of phase-ordering in inhomogeneous systems.
A natural question is then if (and how) the results of this Article can be 
extended to more general situations. For instance,
one might wonder  if a similar picture holds in systems where
dilution is random instead of being, as in this paper, deterministic.
Following the arguments developed in Secs. \ref{scalteo},\ref{scalgraph} 
one realizes that neither the deterministic character, neither
the fractal nature are really determinant. Instead, the fundamental
ingredient is weather the position of minimum energy of interfaces
contain a number $n_c$ of broken bonds which is independent
on $L(t)$ or if such number scales with (some power of)
$L(t)$. This in turn is related to fact that the corresponding graph
does not sustain a ferromagnetic phase ($T_c=0$) or 
it does ($T_c>0$), respectively. Extending this argument to the
case of random dilution we can predict an asymptotic 
logarithmic growth law as
in Eq. (\ref{logl}) for the randomly diluted 
(bond or site) Ising model with a fraction of occupied sites
(or bonds) $p>p_c$, where $p_c$ is the percolation threshold.
Right at $p=p_c$, on the other hand, we expect a power-law growth
as in Eq. (\ref{zt}). Notice that, at variance with the fractal models
studied in this Article, for randomly diluted systems one can 
tune continuously the parameter $p$ which controls the
{\it closeness} to the threshold case $p=p_c$. 
Studies are in progress in order to test these ideas.
Finally, it would be interesting to understand if similar concepts can be
extended to interpret different system, where the inhomogeneous 
character is not due to dilution but to other agents, such as random
coupling constants or spatially varying external fields.

\end{document}